\documentclass[letterpaper, 10 pt, conference]{ieeeconf}  

\usepackage{booktabs}

\IEEEoverridecommandlockouts                              

\usepackage{graphicx}
\usepackage{url}  
\usepackage{pifont}
\usepackage{comment}
\usepackage{amsmath}

\usepackage{lipsum}
\usepackage{amssymb}
\usepackage{multirow}
\usepackage{booktabs}
\usepackage{amssymb}

\title{\LARGE \bf
Environmental Sound Deepfake Detection \\ Using Deep-Learning Framework}

\author{Khoi~Vu*,
        Dat~Tran*,
        Khanh~Do*, 
        Phat~Lam, 
        Vu~Nguyen, 
        Khoa~Nguyen,\\ 
        David~Fischinger, 
        Tin~Nguyen,
        Ian~McLoughlin,
        Son~Le,
        Lam~Pham$\dagger$
\thanks{L. Pham and D. Fischinger are with Austrian Institute of Technology (AIT), Austria}
\thanks{D. Tran and K. Vu are with FPT University, Vietnam}
\thanks{S. Le is with Ton Duc Thang University, Vietnam}
\thanks{P. Lam, T. Nguyenm, and K. Do are with HCM University of Technology, Vietnam}
\thanks{K. Nguyen is with TDT University, Vietnam}
\thanks{I. McLoughlin is with Singapore Institute of Technology, Singapore}
\thanks{($\dagger$) Corresponding author.}
\thanks{(*) Main and equal contribution into the paper.}
}

\begin{document}

\maketitle
\thispagestyle{empty}
\pagestyle{empty}

\begin{abstract}
In this paper, we propose a deep-learning framework for Environmental Sound Deepfake Detection (ESDD) -- the task of identifying whether the sound scene and sound event in an input audio recording is fake or real. 
To this end, we first conducted extensive experiments to explore how individual spectrograms, a wide range of network architectures, and pre-trained models affect the performance of an ESDD model. 
The experimental results on the benchmark datasets of EnvSDD indicate that detecting deepfake audio of sound scene and detecting deepfake audio of sound event should be considered as individual tasks. 
We also indicate that the approach of finetuning a pre-trained model is more effective compared with training a model from scratch for ESDD task. 
Eventually, our best model, which fine-tuned the pre-trained BEATs model with the proposed two-phase training strategy, achieved the Accuracy of 0.98, F1 score of 0.95, AuC score of 0.99 on the Test subset of EnvSDD dataset.
Our best model also achieved the Accuracy of 0.86, F1 score of 0.80, and AuC score of 0.93 when we conducted cross-dataset evaluation on the ESDD-Challenge-TestSet dataset.

\indent \textit{Items}--- Spectrogram, Deepfake, Audio Embedding, Finetuning, Pre-trained Model.
\end{abstract}

\section{INTRODUCTION}
\label{intro}

AI-based generation systems, which are now able to create very realistic audio, have been applied for a wide range of applications such as media production, AI-based music composer, audio-based chatbot, etc.
However, these generation systems also pose potential risks when generated audio, referred to as fake audio or deepfake audio, is used for criminal purposes.
While the research community on audio mainly focuses on detecting deepfakes in speech or in singing voice, deepfake audio of environmental sounds has recently gained the attention. 
Indeed, the first four audio datasets, which were proposed for the ESDD task, have been recently published: FoleySound~\cite{data_03} (2023), SceneFake~\cite{data_02} (2024), EnvSDD~\cite{data_01}(2025), and ESDD-Challenge-TestSet~\cite{source_5}.
Among these datasets, only EnvSDD~\cite{data_01} comprises fake audio for both sound scene and sound event. This dataset also presents the largest number of audio recordings and total recording time.
\begin{figure}[t]
    \centering
      \scalebox{0.8} {
    \includegraphics[width=1.0\linewidth]{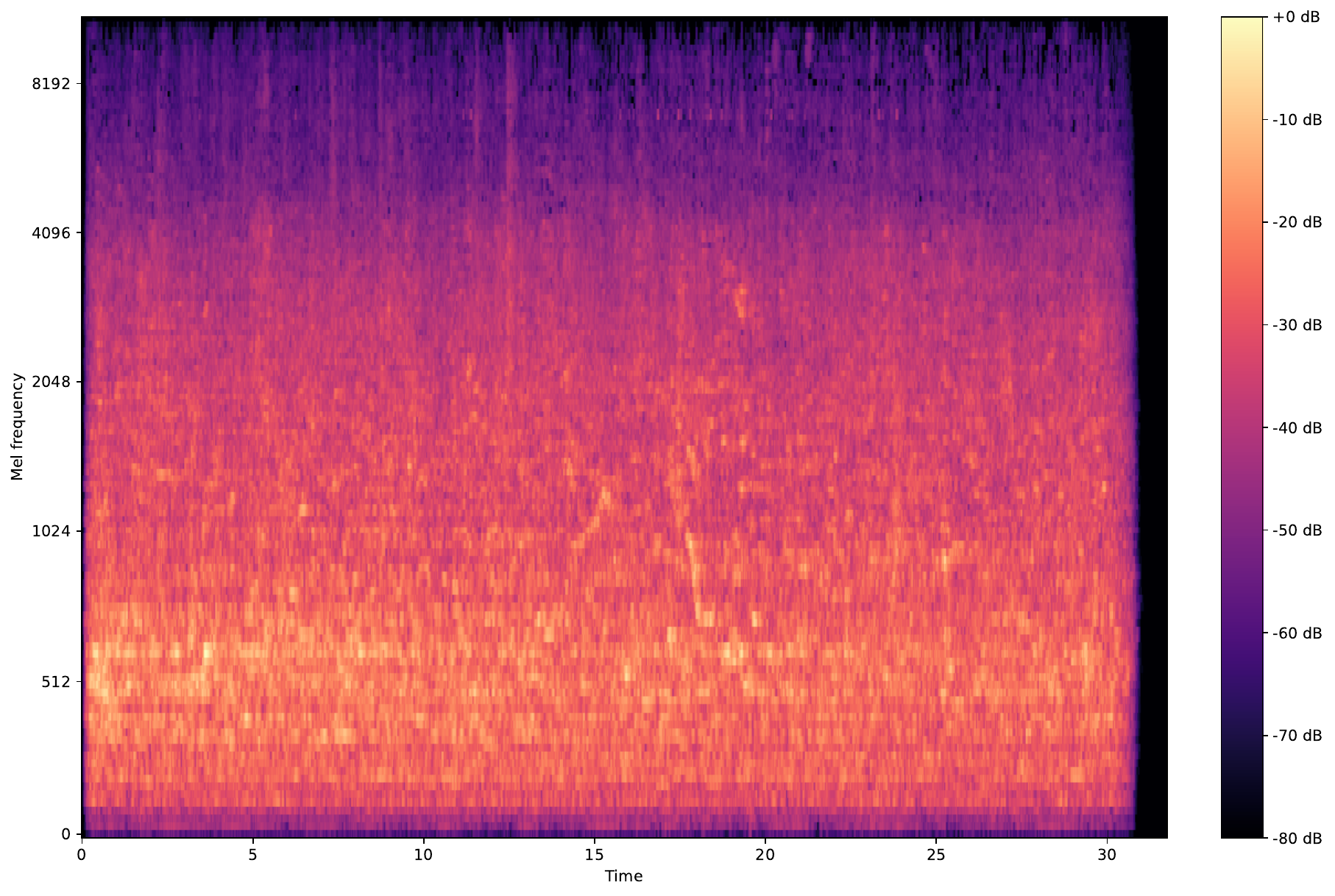}
    }
    \caption{Mel spectrogram of sound scene `in shopping mall'}    
    \label{fig:spec01}
\end{figure}
\begin{figure}[t]
    \centering
      \scalebox{0.8} {
    \includegraphics[width=1.0\linewidth]{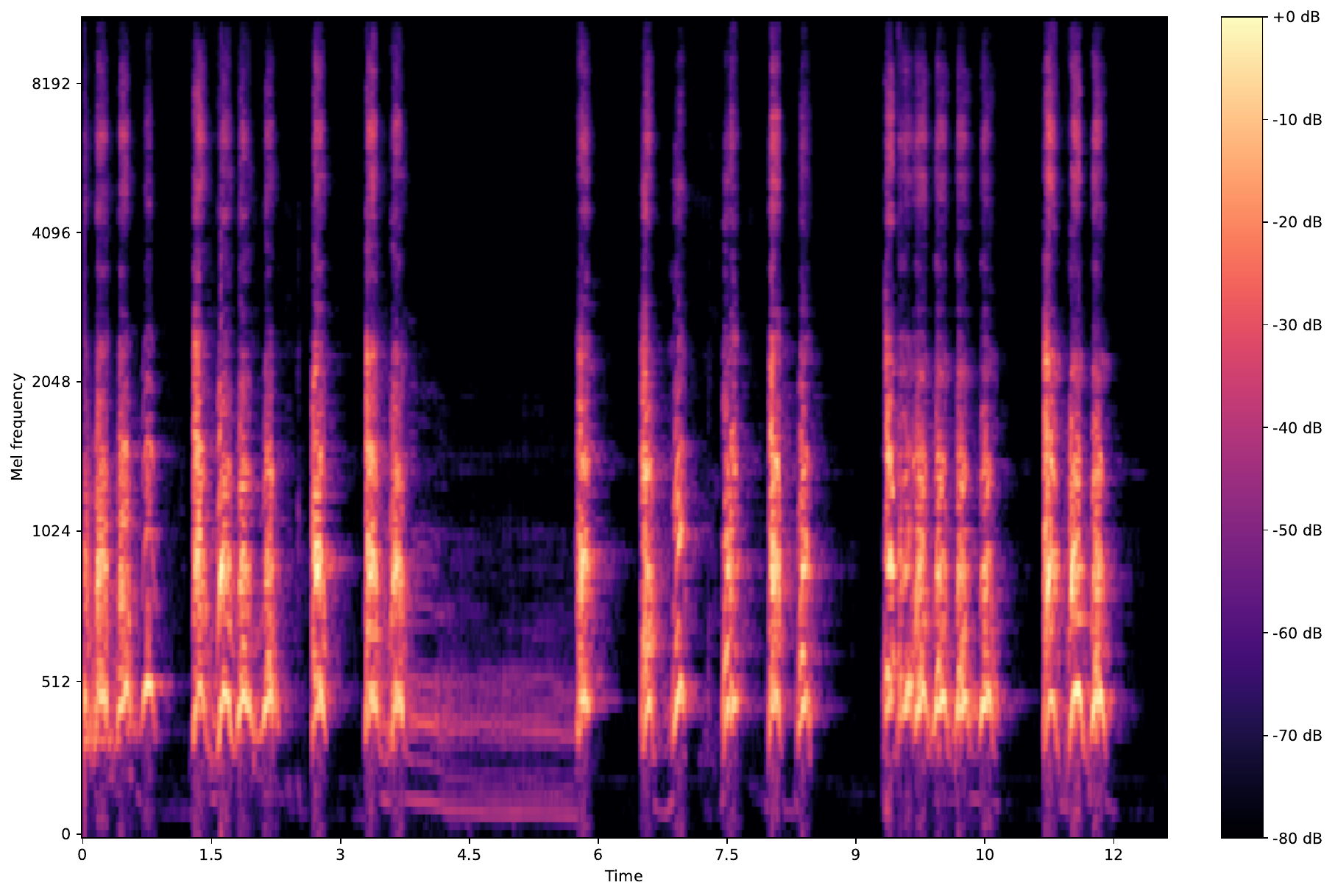}
    }
    \caption{Mel spectrogram of sound event `dog barking'}    
    \label{fig:spec02}
\end{figure}
Regarding the environmental sound, it comprises two main audio types: sound events (e.g., the sound of dog bagging, the sound of siren, etc.) and sound scene (e.g., sound on a bus, sound in a park, etc.).
In comparison to speech, both sound scene and sound event present unstructured in form.
Sound scenes are closely similar to noise in certain contexts and normally are presented in long-time durations.
Meanwhile, there are various sound events in real life which spread in a wide range of frequency bands and are presented in sort-time durations.
For an instance, when looking at the Mel spectrograms of `\textit{in shopping mall}' sound scene and `\textit{dog barking}' sound event in Fig.~\ref{fig:spec01} and Fig.~\ref{fig:spec02}, it can be seen that the spectral power distribution of these two audio types are very different.
In particular, spectral power of the `\textit{in shopping mall}' spreads on all time recording and widely distributes on frequency bands.
Meanwhile, sound event of `\textit{dog barking}' locates on certain on-set time \& off-set time and the significantly spectral power spreads on certain frequency bands.
This leads to raise a research question that whether deepfake detection of sound scene and sound event should be treated as individual tasks.

\begin{figure*}[t]
    \centering
      \scalebox{0.8} {
    \includegraphics[width=1.0\linewidth]{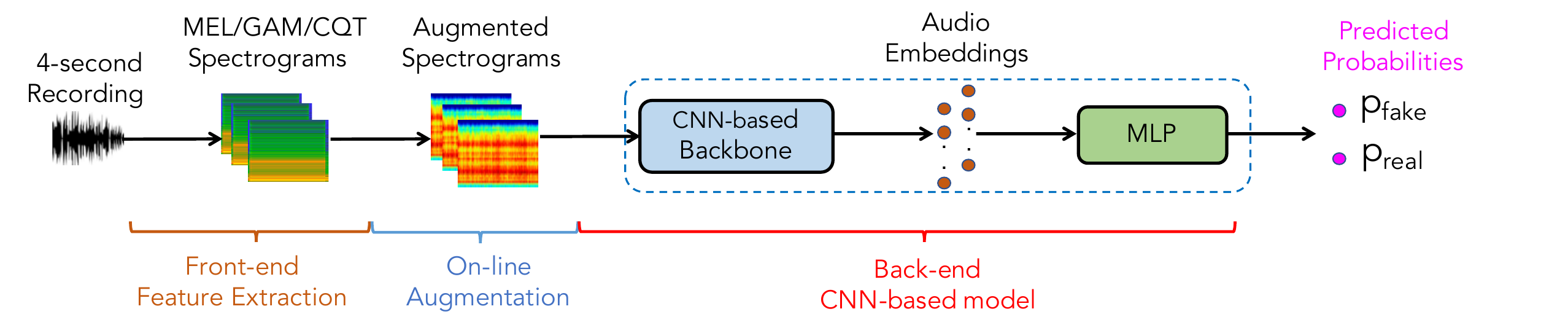}
    }
    \caption{The high-level architecture of the proposed deep-learning framework}    
    \label{fig:high-level-arc}
\end{figure*}

As environmental sounds present different and diverse acoustic characteristics compared with human speech, proposed methods for detecting deepfake of human speech or singing may be less effective and need to be evaluated on environmental sounds.
However, a few of papers~\cite{paper_01, paper_02, envsdd_challenge} proposed for the ESDD task have been published, especially focusing on deepfake audio of sound event.
Recently, the summary of EnvSDD challenge in~\cite{envsdd_challenge} indicates that leveraging the pre-trained models such as BEATs~\cite{beats} is effective to achieve the high-performance ESDD model.
This poses another research question that training a model from scratch or leveraging a pre-trained model is more effective. 
Additionally, a pre-trained model used for ESDD task should be trained on a large-scale dataset of environmental sound such as the pre-trained BEATs~\cite{beats} on AudioSet~\cite{audioset}, or this should be trained on a human speech dataset such as Wav2Vec~\cite{wav2vec20} on LibriSpeech~\cite{MultiLibrispeech}.

As ESDD task has been drawn attention and there are research questions recently mentioned, these inspire us to provide a comprehensive evaluation for ESDD task in this paper. In particular, our main contributions are:
\begin{itemize}
    \item We conducted extensive experiments to evaluate the role of different input spectrograms and various deep neural network architectures on sound event and sound scene independently within the ESDD task. Given the experimental results, we indicate that (1) the approach of leveraging pre-trained model is more effective rather then training a model from scratch for ESDD task, (2) deepfake detection of sound event and sound scene should be treated as two tasks independently, (3) a deepfake detection model trained on sound event can achieve a high performance on sound scene but not vice versa. 

    \item We proposed a deep-learning based model for ESDD task in which the pre-trained BEATs model was finetuned with our novel two-phase training strategy. We evaluated our proposed model on the benchmark dataset of EnvSDD~\cite{data_01}. We also present a cross-dataset evaluation with the ESDD-Challenge-TestSet~\cite{source_5} dataset to verify the model generality (Notably, none of paper has conducted cross-dataset evaluation with respect to the ESDD task). From the experimental results, we indicate that (1) with respect to ESDD task, leveraging a pre-trained model trained on environmental sound  is more effective rather than one trained on human speech dataset, (2) our proposed training strategy with a focus on real distribution presents an effective approach to achieve a general ESDD model.

\end{itemize}

\section{Proposed Deep-Learning Framework}

\subsection{The High-Level Architecture of The proposed Deep-learning Framework}
\label{highlevel}

We first propose a deep-learning based framework for Environmental Sound Deepfake Detection (ESDD). 
The high-level architecture of our proposed framework is described in Fig.~\ref{fig:high-level-arc}.
In particular, the audio is first transformed into spectrograms. 
To deal with the unbalanced issue between fake and real audios, Mixup~\cite{mixup1, mixup2}, is applied as the online data augmentation. Then, original and Mixup spectrograms are fed into a Convolutional-based (CNN-based) network architecture, referred to as the backbone, to extract the audio embedding (i.e., an audio embedding is a vector-based presentation). 
Finally, the audio embedding is explored by a multilayer perception (MLP) with some dense layers to classify audio embeddings into fake or real categories.


\begin{figure*}[t]
    \centering
      \scalebox{0.85} {
    \includegraphics[width=1.0\linewidth]{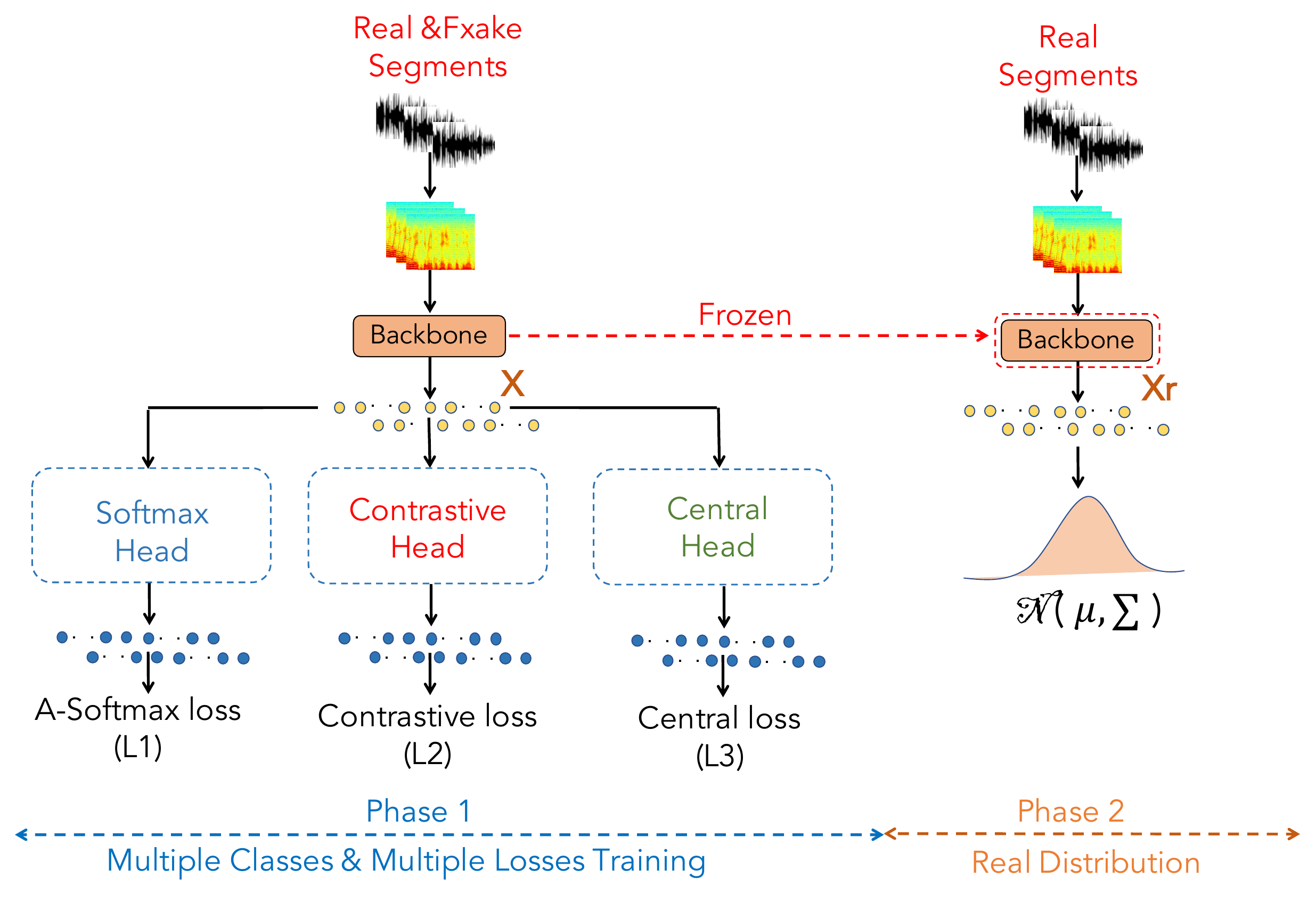}
    }
    \caption{The proposed two-phase training strategy}    
    \label{fig:training_strategy}
\end{figure*}

\subsection{Multiple Input Spectrogram Evaluation}

As each type of spectrogram presents distinct acoustic feature and ensemble of multiple spectrogram inputs is effective to enhance the audio-relevant task performance (i.e., The ensemble of input spectrograms proved effectively with respect to audio tasks of Acoustic Scene Classification and Acoustic Event Detection~\cite{lam_01, lam_02, lam_03, lam_04}, this inspires us to evaluate the role of different spectrograms for the task of ESDD in this paper. 
Therefore, we conduct experiments on three different spectrograms of  Constant Q Transform (CQT), (MEL), and Gammatone (GAM). 
We also evaluate if an ensemble of these spectrograms is possible to enhance the ESDD task performance. 

\subsection{Deep Neural Network Backbone Evaluation}

To achieve the most effective network architecture, we evaluate a wide range of network architectures such as VGG16, MobileNetV2, ConvNeXt-Tiny, NASNet-Large, ResNet50, Inception-V3, EfficientNetB1, DenseNet161, Xception.
Additionally, we evaluate different pre-trained models such as BEATs~\cite{beats}, Whisper~\cite{whisper}, wavLM~\cite{wavlm}, Wave2Vec~\cite{wav2vec20} which were trained on the large-scale datasets. 
Notably, BEATs model~\cite{beats} was trained on sound event dataset of AudioSet~\cite{audioset}. 
Meanwhile, Whisper~\cite{whisper}, WavLM~\cite{wavlm}, and Wave2Vec~\cite{wav2vec20} were trained on speech datasets.
These network architectures are used as the backbone and then are connected with the MLP component as mentioned in Section~\ref{highlevel} and Fig.~\ref{fig:high-level-arc}.

\subsection{The Two-phase Training Strategy}
\label{sec_fine}
\begin{table*}[t]
 \caption{The number of 4-second real/fake audio in EnvSDD dataset \\ (the development subset is upper and the testing subset is lower)}
  \centering
  \scalebox{0.75} {
  \begin{tabular}{|c| l  | c  | c  c c c c c c |}
    \hline
    &Data Resource  & Real Audio   & \multicolumn{7}{c|}{Fake Audio} \\
                   & &       & ATA-Audioldm1 &TTA-audiogen &TTA-audioldm1 &TTA-audioldm2 &ATAAudioldm2 &TTA-Audiolcm &TTA-Tangoflux\\
    \hline   
    \hline
    &(1) TUTASC2019Dev (Acoustic Scene)   &28530 &28530 &28530 &28530 &28530 &None &None &None \\
    &(2) TUTSED2016Dev (Acoustic Event)  &1048 &1048 &1048 &1048 &1048 &None&None&None\\
    Development&(2) TUTSED2016Eval (Acoustic Event) &478 &478 &478 &478 &478 &None&None&None\\
    Subset & (3) TUTSED2017Dev (Acoustic Event)  &1235 &1235 &1235 &1235 &1235 &None&None&None\\
    &(3) TUTSED2017Eval (Acoustic Event) &391 &391 &391 &391 &391 &None&None&None\\ 
    &(4) UrbanSound8K (Acoustic Event)   &4071 &4071 &4071 &4071 &4071 &None&None&None\\
        \hline
        \hline              
     &(1) TUTASC2019Dev (Acoustic Scene)   &3170 &3170 &3170 &3170 &3170 &3170 &3170 &3170 \\
     &(2) TUTSED2016Dev (Acoustic Event)   &116 &116 &116 &116 &116 &116 &116 &116\\
    &(2) TUTSED2016Eval (Acoustic Event)  &53 &53 &53 &53 &53 &53 &53 &53\\
    Test& (3) TUTSED2017Dev (Acoustic Event)   &137 &137 &137 &137 &137 &137 &137 &137 \\
    Subset& (3) TUTSED2017Eval (Acoustic Event)  &43 &43 &43 &43 &43 &43 &43 &43 \\ 
    &(4) UrbanSound8K (Acoustic Event)    &452 &452 &452 &452 &452 &452 &452 &452\\
    & (5) DCASE2023-Task7 (Acoustic Event) &500 &500 &500 &500 &500 &500 &500 &500 \\
    & (6) Clotho (Acoustic Event)          &500 &500 &500 &500 &500 &500 &500 &500 \\

        \hline
  \end{tabular}
  }
  \label{tab:dataset}
\end{table*}
In our previous work~\cite{novel_train}, we proposed a novel training strategy which proves effective to separate real and fake distribution for deepfake speech detection.
This inspires us to apply a variant of this training strategy for the ESDD task in this paper.
Particularly, the training strategy is modified and described in Fig.~\ref{fig:training_strategy} with two phases of training processes.

In the first phase, we train the model with multiple heads accompanying with multiple loss functions. 
In particular, batch of audios  go through the backbone to obtain audio embedding $\mathbf{X}$. 
Then, multiple heads, each of which presents one MLP architecture as shown in Fig.~\ref{fig:high-level-arc}, are applied to explore the audio embedding $\textbf{X}$.
The first loss function, A-Softmax loss, is for multiple-class classification (i.e., we consider each fake generator as a category). 
The second loss, Contrastive loss~\cite{Contrastive_loss}, is used to separate real and fake distribution. 
The final loss, Central loss, is used to condense the distribution of real audio. 
The combination of three loss functions is effective to separate fake and real distribution.

In the second phase, we feed real audios into the backbone, obtain real audio embeddings $\mathbf{X_r}$.
Given real audio embeddings, we compute and achieve the central point of the  distribution of real audio embeddings $\mathcal{N(\mathbf{X_r})}$. 
Regarding the reference process, a testing audio is fed into the backbone to obtain a testing audio embedding $\mathbf{x_t}$. 
The testing audio embedding is compared with the central point of the real distribution $\mathcal{N(\mathbf{X_r})}$ using cosine similarity to decide whether the testing audio is fake or real.

The idea of using the central point of the real audio distribution to decide an unseen audio as fake or real is inspired by: (1) There are a lot of real audio events (i.e., AudioSet~\cite{audioset} presents the most diverse and largest datasets for audio events) while lacking of fake audio events with a limitation of environmental sound fake generators. Given diverse real audio events, if we successfully train a model that makes the distribution of real audio embedding more condense and well separating, using the distribution of real audio for detecting unseen fake audio is effective; (2) A model normally presents the overfitting issue on the training data. Therefore, fake and real audio from the training data should be separated well. However, if unseen fake audio locates at the boundary between training fake and real audio, this leads the incorrect detection if basing on the boundary threshold. Instead, if using the distribution of real audio, which is condense and far away the boundary threshold, this helps to avoid the mis-classification regarding the unseen fake audio at the boundary threshold.

\section{Evaluation Datasets and Experimental Settings}
\subsection{Evaluation Datasets}
As mentioned in Section~\ref{intro}, only four datasets of  FoleySound~\cite{data_03} (2023), SceneFake~\cite{data_02} (2024), EnvSDD~\cite{data_01}(2025), and ESDD-Challenge-TestSet~\cite{source_5} have been proposed for ESDD task. 
While FoleySound~\cite{data_03}, ESDD-Challenge-TestSet~\cite{source_5} are only for sound event and SceneFake~\cite{data_02} is only for sound scene, EnvSDD~\cite{data_01} includes both sound event and sound scene. 
Therefore, we first use EnvSDD~\cite{data_01} dataset to conduct extensive experiments to analyze and achieve the best model. We use ESDD-Challenge-TestSet~\cite{source_5} to verify the generality of our best model by cross-dataset evaluation.
As the scope of this paper focus on sound scene and sound event without human speech, we do not use SceneFake~\cite{data_02} dataset in which human speech is added. We also do not use FoleySound~\cite{data_03} in this paper as this dataset is from DCASE2023-Task7 and is included in EnvSDD dataset~\cite{data_01}.

\textbf{EnvSDD dataset~\cite{data_01}:} As shown in Table~\ref{tab:dataset}, EnvSDD dataset comprises two subsets of the development subset and the test subset.
The real audio in the EnvSDD development subset are from four main resources: DCASE Challenge 2019 Task 1 (TUTASC2019Dev)~\cite{source_1} (e.g., audio scene), DCASE Challenge 2016 Task 3 (TUTSED2016Dev \& TUTSED2016Eval)~\cite{source_2} (e.g., audio event), DCASE Challenge 2017 Task 3 (TUTSED2017Dev \& TUTSED2017Eval)~\cite{source_3} (e.g., audio event), and finally UrbanSound8K~\cite{source_4} (e.g., audio event).
For each data resource, the number of 4-second real audio recordings are collected and shown in the second column of Table~\ref{tab:dataset}. 
Then, four different generated-AI systems (e.g., ATA-Audioldm1, TTA-audiogen,  TTA-audioldm1, and TTA-audioldm2), using two main techniques of audio-to-audio (ATA) and text-to-audio (TTA), are applied to generate fake audio from real audio.
Meanwhile, the real audio in the EnvSDD test subset not only comprise four resources as mentioned in the development set, but also includes DCASE2023Task7~\cite{dcase20213t7} (e.g., audio event) and Clotho~\cite{data_clotho} (e.g., audio event).
Additionally, three different generated-AI systems of ATA-Audioldm2, TAT-Audiolcm, and TAT-Tangoflux are added to generate fake audio in the test subset. 
Given data in Table~\ref{tab:dataset}, there are two main concerns. First, the EnvSDD dataset presents unbalanced with the dominant TUTASC2019Dev. Second, while TUTASC2019Dev presents sound scene recordings, other data resources collect sound events. 

\textbf{ESDD-Challenge-TestSet~\cite{source_5}:} To evaluate the generality of proposed models, we conduct the cross-dataset evaluation with ESDD-Challenge-TestSet.
This dataset comprises 1994 4-second bonafide audio and 7981 fake audio.
The bonafide audio is from VGG-Sound~\cite{source_6} which presents sound events.
The fake audio is generated from two generators, referred to as `diff\_foley' and `foleycrafter'. 

\subsection{Experimental Settings}

\textbf{Proposed Testing Scenario:} We follow the paper~\cite{data_01}, use the development subset of EnvSDD to train the proposed models and evaluate the proposed models on the test subset of EnvSDD.
Given the data splitting, we first propose three test cases. 

The first test case (\textbf{Test-Case-1}) is for the task of Sound Scene Fake Detection (SSFD) in which only sound scene recordings (e.g., only real and fake audio from TUTASC2019) are used for training and testing.

Meanwhile, the second test case  (\textbf{Test-Case-2}) is for Sound Event Fake Detection (SEFD). 
Similarly, only audio recordings containing fake and real audio events (e.g., EnvSDD dataset with the exception of TUTASC2019) are used for training and testing.

In the final test case (\textbf{Test-Case-3}) , we cross-test between two tasks of SSFD and SEFD.
In particular, the model for SSFD is used to test on SEFD task and vice versa.

From the experimental results from three test cases, we propose the best model for individual tasks of SSFD and SEFD.
Given the best model, we finally conduct the cross-dataset evaluation on the datasets of ESDD-Challenge-TestSet~\cite{source_5} to evaluate the model generality.

\textbf{Settings:}
All proposed models in the paper are implemented with Pytorch framework. 
All experiments in this paper are run with GPU Titan 23 GB . 
Adam algorithm~\cite{Adam} is used for the optimization.

\textbf{Evaluation Metrics:} In this paper, we use Accuracy, F1 score, and AuC score as main metrics to compare performance among models. We also present the confusion matrix on cross-dataset evaluation. We only report EER score when we compare with the state-of-the-art models.
On other words, we do not treat EER score as the main metric as an EER score achieved from testing on a dataset always come together with a threshold. By testing on different datasets, the corresponding thresholds may be different. This does not reflect the real scenario for the inference process in which only one threshold is decided for predicting an unseen audio.

\begin{table}[t]
 \caption{\textbf{Test-Case-1}: Acoustic Scene Fake Detection (ASFD) on EnvSDD dataset}
  \centering
  \scalebox{0.9} {
  \begin{tabular}{|c | c| c |c |c|}
    \hline
    \textbf{Spectrograms}   & \textbf{Models}       &\textbf{Acc.} &\textbf{F1} &\textbf{AuC} \\    
    \hline
                 \hline
    MEL           &EfficientNetB1+MLP           &0.88  &0.47   &0.69 \\
    CQT           &EfficientNetB1+MLP           &0.91  &0.83   &0.96 \\
    GAM           &EfficientNetB1+MLP           &0.95  &0.89   &0.99 \\ 
    All Spec. (ensemble)   &EfficientNetB1+MLP           &0.96  &0.90   &0.99 \\
    \hline
    GAM           &EfficientNetB1+MLP        & 0.95  &0.89 & \textbf{0.99} \\ 
    GAM           &DenseNet161+MLP           & 0.92 & 0.85 & \textbf{0.99} \\
    GAM           &Inception-V3+MLP          & 0.91 & 0.84 & 0.98 \\
    GAM           &ResNet50+MLP              & 0.89 & 0.81 & 0.96 \\
    GAM           &VGG16+MLP                   & \textbf{0.97} & 0.89 & \textbf{0.99} \\
    GAM           &MobileNetV2+MLP              & 0.93 & 0.79 & \textbf{0.99} \\
    GAM           &ConvNeXt-Tiny+MLP              & 0.96 & 0.85 & \textbf{0.99} \\
    GAM           &Xception+MLP                   & 0.96 & 0.86 & \textbf{0.99} \\
    GAM           & NASNet-Large+MLP&   0.94&  0.80&  0.97\\
    \hline
    Raw Audio   &BEATs-Emb+MLP &0.95 &\textbf{0.94} &0.96 \\
    \hline
  \end{tabular}
  }
  \label{tab:scene}
\end{table}

\begin{table}[t]
 \caption{\textbf{Test-Case-2}: Acoustic Event Fake Detection (AEFD) on EnvSDD dataset}
  \centering
  \scalebox{0.9} {
  \begin{tabular}{|c|  c| c| c| c| }
    \hline
    \textbf{Spectrograms}   & \textbf{Models}       &\textbf{Acc.} &\textbf{F1} &\textbf{AuC} \\    
    \hline
                 \hline
    MEL           &EfficientNetB1+MLP          & 0.36 & 0.34 & 0.59 \\
    CQT           &EfficientNetB1+MLP          & 0.69 & 0.50 & 0.51\\
    GAM           &EfficientNetB1+MLP          & 0.74 & 0.62 & 0.79 \\ 
    All Spec. (ensemble)   &EfficientNetB1+MLP          & 0.78 & 0.67 & 0.81 \\
    \hline
    GAM           &EfficientNetB1+MLP          & 0.74 & 0.62 & 0.79 \\ 
    GAM           &DenseNet161+MLP           & 0.83 & 0.67 & 0.77 \\
    GAM           &Inception-V3+MLP           & 0.81  &0.66  &0.78 \\
    GAM           &ResNet50+MLP              & 0.82  &0.66  &0.78 \\
    GAM           &VGG16+MLP                   &   0.85&  0.50&  0.83\\
    GAM           &MobileNetV2+MLP                   &   0.79&  0.41&  0.77\\
    GAM           &ConvNeXt-Tiny+MLP                   &   0.85&  0.49&  0.84\\
    GAM           &Xception+MLP                   &   0.84&  0.46&  0.80\\
    GAM           &NASNet-Large+MLP                   &   0.84&  0.46&  0.82\\
    \hline
    Raw Audio   &BEATs-Emb+MLP &\textbf{0.96} &\textbf{0.97} &\textbf{0.95} \\
     \hline

  \end{tabular}
  }
  \label{tab:event}
\end{table}

\begin{table}[t]
 \caption{\textbf{Test-Case-3}: Cross-Test Evaluation Between ASFD and AEFD on EnvSDD dataset}
  \centering
  \scalebox{0.8} {
  \begin{tabular}{|c | c |c| c| c| c|}
    \hline
    \textbf{Models}   & \textbf{Train} &\textbf{Test}           &\textbf{Acc.} &\textbf{F1} &\textbf{AuC} \\ 
    \hline
                 \hline
    GAM+EfficientNetB1+MLP   &ASFD (scene)&AEFD (event) & 0.87 & 0.60 & 0.61 \\
    GAM+InceptionV3+MLP      &ASFD (scene)&AEFD (event) & 0.88 & 0.60 & 0.66 \\
    GAM+ResNet50+MLP         &ASFD (scene)&AEFD (event) & 0.87 & 0.59 & 0.66 \\
    GAM+DenseNet161+MLP      &ASFD (scene)&AEFD (event) & 0.88 & 0.60 & 0.68 \\
    BEATs-Emb+MLP   &ASFD (scene)&AEFD (event) & \textbf{0.92} & \textbf{0.89} & \textbf{0.90} \\
    \hline
    GAM+EfficientNetB1+MLP   &AEFD (event)&ASFD (scene) & 0.74 & 0.63 & 0.84 \\
    GAM+InceptionV3+MLP      &AEFD (event)&ASFD (scene) & 0.85 & 0.63 & 0.78 \\
    GAM+ResNet50+MLP         &AEFD (event)&ASFD (scene) & 0.84 & 0.63 & 0.78 \\
    GAM-DenseNet161+MLP      &AEFD (event)&ASFD (scene) & 0.84 & 0.63 & 0.80 \\
    BEATs-Emb+MLP   &AEFD(event)&ASFD(scene) & \textbf{0.93} & \textbf{0.92} & \textbf{0.94} \\
 \hline
  \end{tabular}
  }
  \label{tab:cross}
\end{table}

\begin{table}[t]
 \caption{Evaluate BEATs-finetune+MulHead with the two-phase training strategy on EnvSDD dataset}
  \centering
  \scalebox{0.8} {
  \begin{tabular}{|c|  c |c |c |c |c|}
    \hline
    \textbf{Models}   & \textbf{Train} &\textbf{Test}           &\textbf{Acc.} &\textbf{F1} &\textbf{AuC} \\ 
    \hline
             \hline
    BEATs-finetune+MulHead   &ASFD (scene) &ASFD (scene) &0.99  &0.99 &0.99  \\
    BEATs-finetune+MulHead   &ASFD (scene) &AEFD (event) &0.95 &0.89 &0.9  \\
    \hline
    BEATs-finetune+MulHead   &AEFD (event) &AEFD (event) &0.94 &0.88 &0.98  \\
    BEATs-finetune+MulHead   &AEFD (event) &ASFD (scene) &0.96 &0.92 &0.99 \\
    \hline
    BEATs-finetune+MulHead   &ASFD\&AEFD  &ASFD (scene) & 0,98 & 0.96 & 0.99 \\
     &(scene\&event) & &&&\\
    BEATs-finetune+MulHead   &ASFD\&AEFD  &AEFD (event) & 0.96 & 0,92 & 0.99 \\
     &(scene\&event) & &&&\\
     BEATs-finetune+MulHead   &ASFD\&AEFD  &ASFD\&AEFD & \textbf{0.98} & \textbf{0.95} & \textbf{0.99} \\
     &(scene\&event) &(scene\&event) &&&\\
 \hline
  \end{tabular}
  }
  \label{tab:finetune01}
\end{table}

\section{Experimental Results and Discussion}

\textbf{Acoustic Scene Fake Detection (ASFD: Test-Case-1)}: In this test case, we evaluate deep-learning based models for ASFD task.
Therefore, we only train and test with audio from TUTASC2019Dev set (e.g., sound scene). 
As Table~\ref{tab:scene} shows, GAM spectrogram outperforms MEL and CQT spectrograms when using the same network architecture with EfficientNetB1 backbone.
When we conduct ensemble of multiple spectrograms, it is effective to enhance the ASFD performance
As GAM presents better performance compared with CQT and MEL, we then evaluate different network architectures with GAM spectrogram input.
Among network architectures trained from scratch, EfficientNetB1, VGG16 and Xception backbones present the high performance compared with the others.
However, although all these models present Acc. and AuC scores larger than 0.95, the F1 scores are lower than 0.90. 
This indicates that incorrect detection significantly occurs on real audio.
In other words, the overfitting occurs on fake audio (i.e., the number of fake audio is significantly larger than the number of real audio in the training subset).
Regarding the approach of leveraging a pre-trained model, we use the pre-trained BEATs model to extract the audio embeddings. Then, the audio embeddings are explored by a MLP network. We referred the network as to BEATs-Emb+MLP.
We achieve a high performance on the pre-trained BEATs model with Acc. score of 0.95, F1 score of 0.94, and AuC score of 0.96.
These nearly equal scores prove that no overfitting on fake audio happens regarding the BEATs-Emb+MLP model. 

\textbf{Acoustic Event Fake Detection (AEFD: Test-Case-2)}: In this test case, we evaluate deep-learning based models for AEFD task.
Therefore, we train and test with all audio, excepted in TUTASC2019Dev set.
All network architectures used in \textbf{Test-Case-1} are re-used in the \textbf{Test-Case-2} for a comparison.
Same as Test-Case-1, results in Table~\ref{tab:event} indicate that GAM spectrogram presents the best performance among the evaluating spectrograms. 
The ensemble of spectrograms does not help to improve the performance significantly.
Regarding network architectures trained from scratch, DenseNet161+MLP presents a balancing performance among Acc., F1, and AuC scores compared with the others. 
Leveraging the pre-trained BEATs model still achieves the best and balancing performance on all metrics.
Again, the approach of using the pre-trained BEATs model avoid the overfitting issue on AEFD task.
\begin{figure}[t]
    \centering
      \scalebox{1.0} {
    \includegraphics[width=1.0\linewidth] {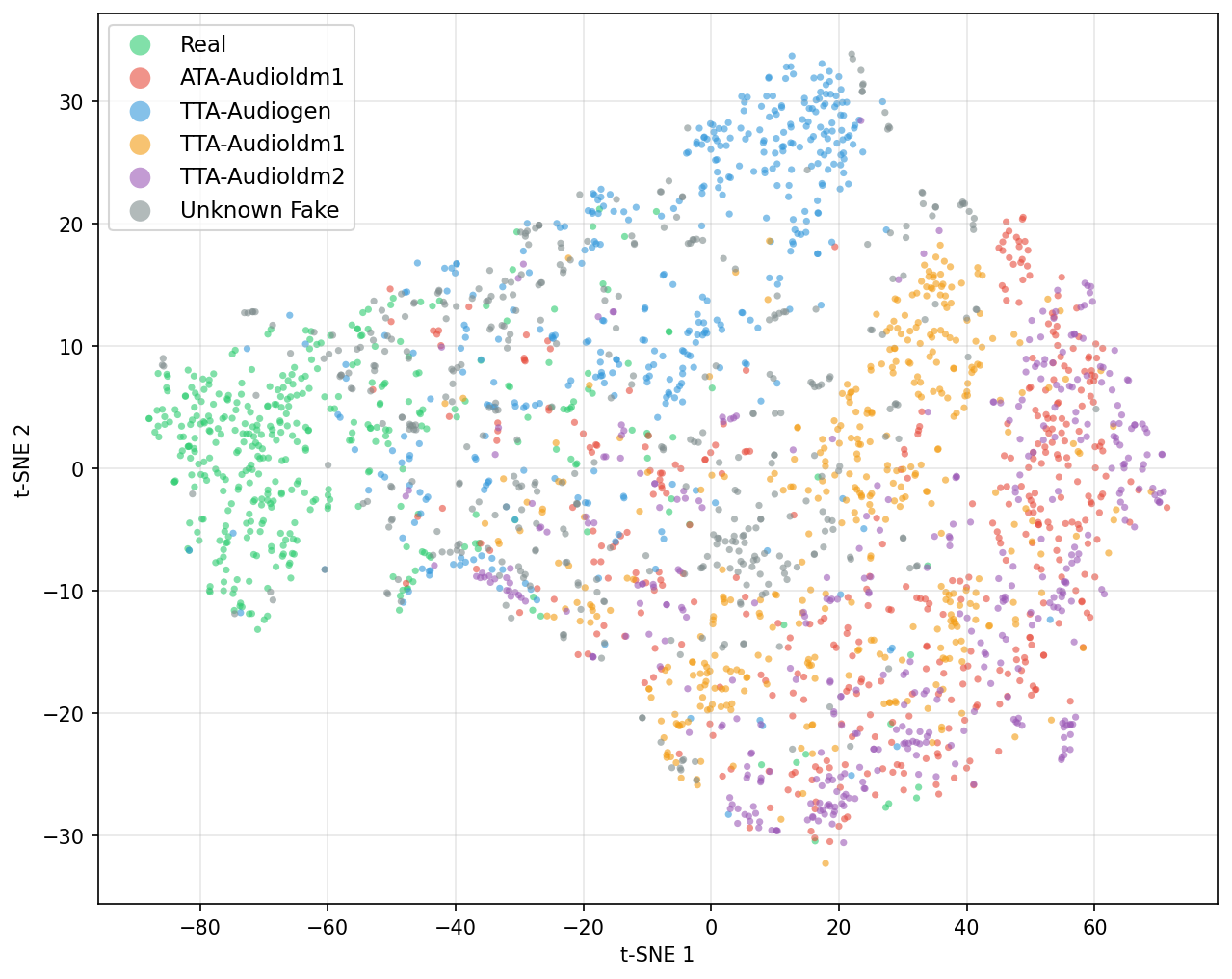}
    }
    \caption{Distribution of \textbf{fake and real audio events} in Test subset of EnvSDD dataset using BEATs-finetune+MulHead model (Unknown fake : ATAAudioldm2, TTA-Audiolcm, TTA-Tangoflux)
    }    
    \label{fig:tsne_01}
\end{figure}

\begin{figure}[t]
    \centering
      \scalebox{1.0} {
    \includegraphics[width=1.0\linewidth]{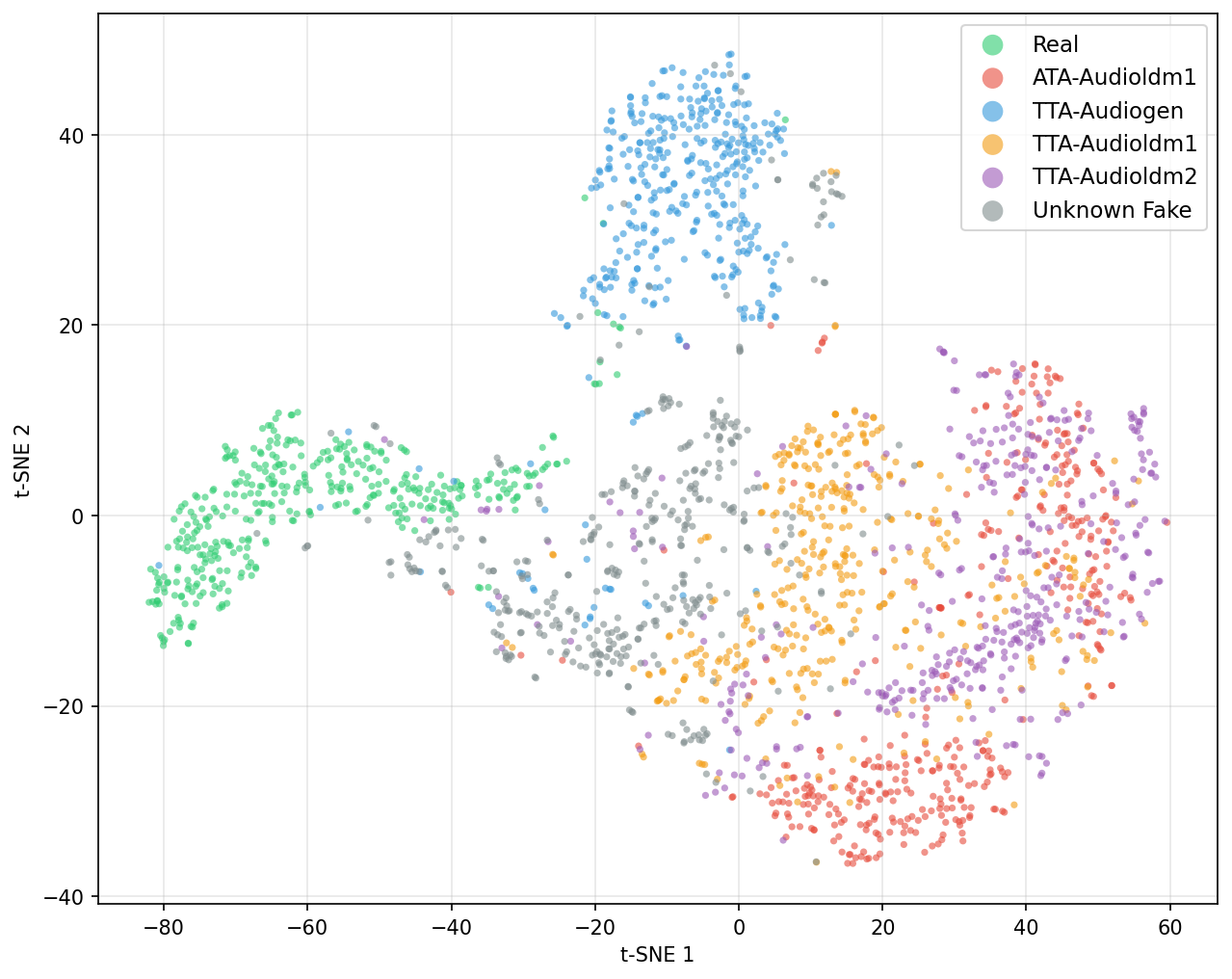}
    }
    \caption{Distribution of \textbf{fake and real audio scenes} in Test subset of EnvSDD dataset using BEATs-finetune+MulHead model (Unknown fake : ATAAudioldm2, TTA-Audiolcm, TTA-Tangoflux)
    }    
    \label{fig:tsne_02}
\end{figure}

\begin{figure}[t]
    \centering
      \scalebox{1.0} {
    \includegraphics[width=1.0\linewidth] {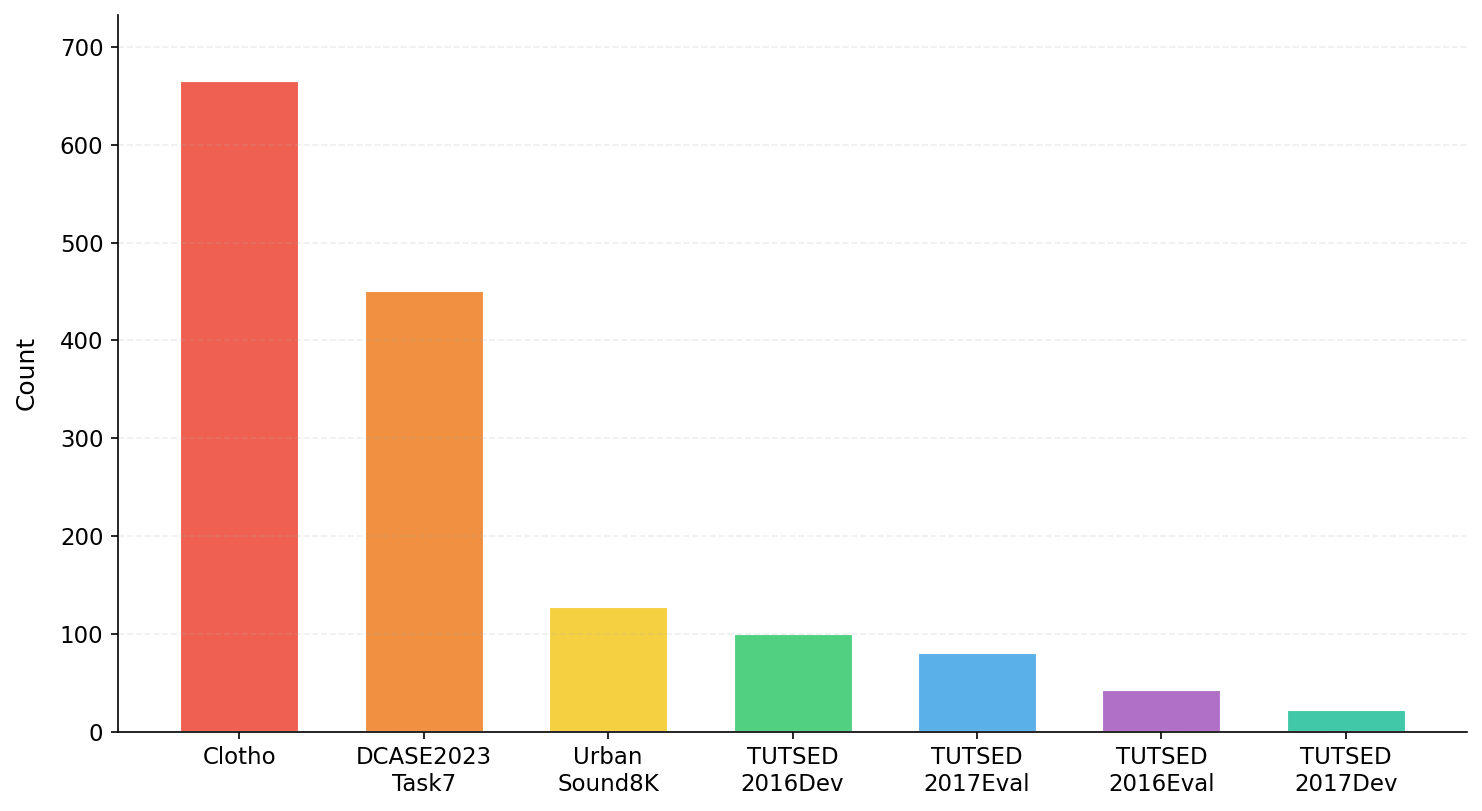}
    }
    \caption{Deepfake Sound Event Incorrect Detection Across Resource Data}    
    \label{fig:inc_01}
\end{figure}

\begin{figure}[t]
    \centering
      \scalebox{1.0} {
    \includegraphics[width=1.0\linewidth] {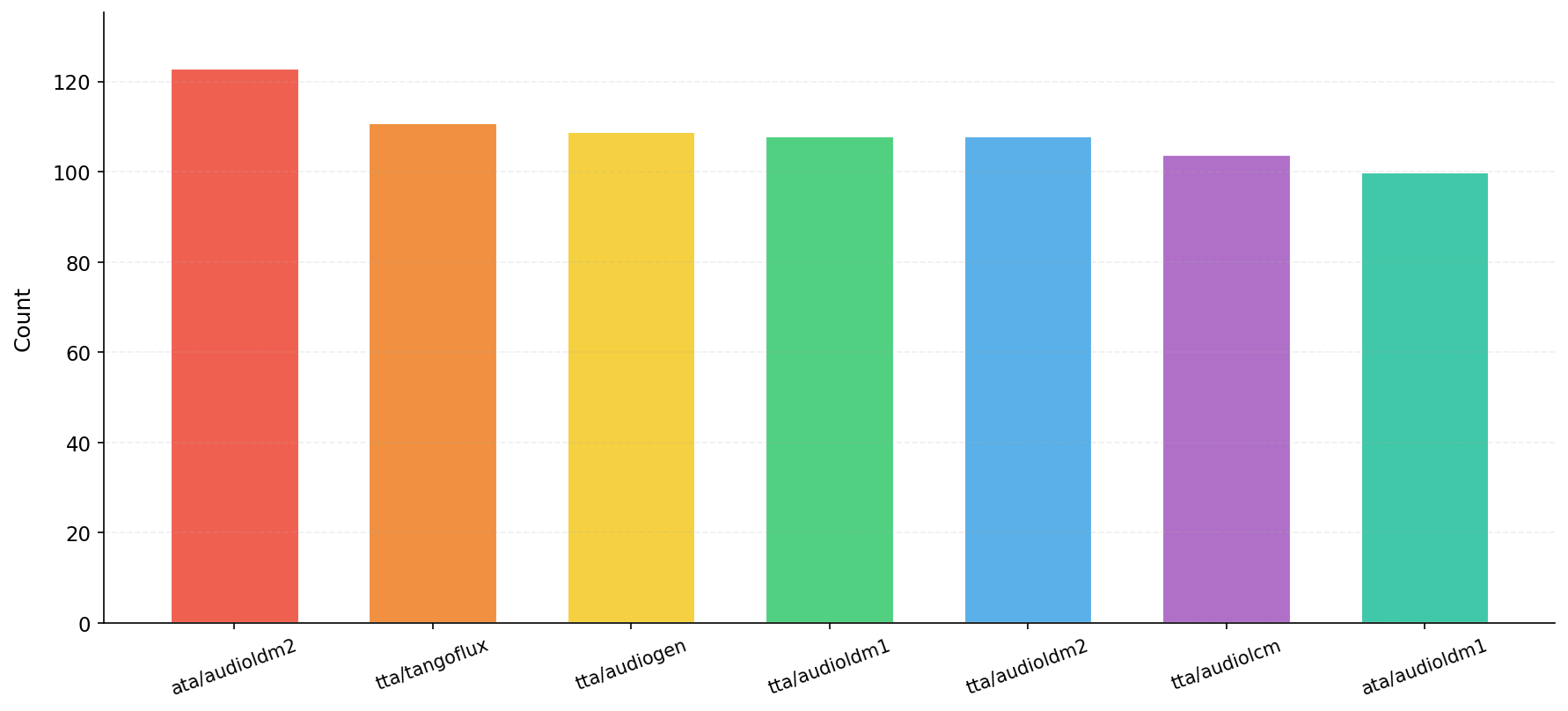}
    }
    \caption{Deepfake Sound Event Incorrect Detection Statistics Across Fake Generators}    
    \label{fig:inc_02}
\end{figure}

\begin{table}[t]
\caption{Performance Comparison with State-of-the-art Models in EER (\%) on EnvSDD dataset. “Seen SD” means “Seen Source Datasets” and “Seen AM” means “Seen Audio Generater”.}
\centering
  \scalebox{0.9} {
\begin{tabular}{|c |c | c| c | c| c|}
\hline
\textbf{Systems} & \textbf{Test Set} & \textbf{Seen}  & \textbf{Seen} & \textbf{TTA} & \textbf{ATA} \\
        &          &\textbf{SD}    & \textbf{AG} & & \\
        \hline
        \hline
 & Test 01 & Yes & Yes                        & 3.1 & 3.4 \\
 & Test 02 & Yes & No                         & 3.8 & 4.6 \\
 Our BEATs-finetune+MulHead& Test 03 & No      & Yes  & 17.2 & 2.1 \\
 (Train on ASFD)& Test 04 & No      & No      & 15.3 & 17.6 \\
 & Average & -- & --                          & \textbf{9.6} & \textbf{17.6}  \\
\hline
\hline
 & Test 01 & Yes & Yes                        & 0.05 & 0.05 \\
 & Test 02 & Yes & No                         & 0.6 & 0.3 \\
 Our BEATs-finetune+MulHead& Test 03 & No      & Yes  & 4.7 & 2.9 \\
 (Train on AEFD)& Test 04 & No      & No      & 13.9 & 1.5 \\
 & Average & -- & --                          & \textbf{4.8} & \textbf{1.18}  \\
\hline
\hline
 & Test 01 & Yes & Yes                       & 0.01 & 0.01 \\
 & Test 02 & Yes & No                        & 0.3  & 0.01 \\
 Our BEATs-finetune+MulHead& Test 03 & No & Yes      & 3.60  & 3.10 \\
 (Train on AEFD\&ASFD)& Test 04 & No & No    & 4.30  & 3.20 \\
 & Average & -- & --                         & \textbf{2.05} & \textbf{1.58} \\
\hline
\hline
                       & Test 01 & Yes & Yes  &0.66  &0.28  \\
                       & Test 02 & Yes & No   &3.70  &0.68  \\
 AASIST~\cite{data_01} & Test 03 & No & Yes   &6.80  &  3.00\\
 (Train on AEFD\&ASFD) & Test 04 & No & No    &17.50 &4.40 \\
                       & Average & -- & --    & \textbf{7.17} & \textbf{2.09} \\
\hline
\hline
                       & Test 01 & Yes & Yes  &0.26  &0.38  \\
                       & Test 02 & Yes & No   &13.04  &26.59  \\
 W2V2+AASIST~\cite{data_01} & Test 03 & No & Yes   &10.60  &13.30  \\
 (Train on AEFD\&ASFD) & Test 04 & No & No    &45.80  &52.40 \\
                       & Average & -- & --    & \textbf{17.43} & \textbf{23.17} \\
\hline
\hline
                       & Test 01 & Yes & Yes      &0.08  &0.03  \\
                       & Test 02 & Yes & No       &1.26  &0.08  \\
 BEATs+AASIST~\cite{data_01} & Test 03 & No & Yes &4.70  &2.20  \\
 (Train on AEFD\&ASFD) & Test 04 & No & No        &17.20 &3.00 \\
                       & Average & -- & --    & \textbf{5.81} & \textbf{1.33} \\
  \hline

\end{tabular}
}
\label{tab:per_cmp}
\end{table}

\begin{table}[t]
 \caption{Cross-dataset evaluation on ESDD-Challenge-TestSet~\cite{source_5}} 
  \centering
  \scalebox{0.8} {
  \begin{tabular}{|c|  c |c |c |c |c|}
    \hline
    \textbf{Models}   & \textbf{Train} &\textbf{Test}           &\textbf{Acc.} &\textbf{F1} &\textbf{AuC} \\ 
    \hline
             \hline
    BEATs-finetune   &EnvSDD train subset  &ESDD-Challenge &   0.53&  0.47&  0.53\\
    MulHead   & (sound scene only) &-TestSet~\cite{source_5} &   &  &  \\
    \hline   
    BEATs-finetune   &EnvSDD train subset  &ESDD-Challenge &   \textbf{0.86}&  \textbf{0.80}&  \textbf{0.93}\\
    MulHead   & (sound event only) &-TestSet~\cite{source_5} &   &  &  \\
    \hline   
    BEATs-finetune   &EnvSDD train subset  &ESDD-Challenge &   0.66&0.60& 0.76\\
    MulHead   & (sound scene \& event) &-TestSet~\cite{source_5} &   &  &  \\ 
    \hline
    
  \end{tabular}
  }
  \label{tab:finetune02}
\end{table}

\textbf{Cross-Test Between ASFD and AEFD (Test-Case-3)}:
Given high-performance models from \textbf{Test-Case-1} and \textbf{Test-Case-2}, we evaluate the cross-test between two tasks of ASFD and AEFD in this \textbf{Test-Case-3}.
In other words, selected models trained on ASFD (sound scene) is test on AEFD (sound event) and vice versa.
As the results are shown in Table~\ref{tab:cross}, it indicates that training a model on ASFD (sound scene) and testing on AEFD (sound event) is not effective. 
This is similar when we train a model from scratch on AEFD (sound event) and testing on ASFD (sound scene event). 
However, training on AEFD (sound event) and testing on ASFD (sound scene) using the pre-trained BEATs model (BEATs-Emb-MLP) is very competitive compared with training and testing on ASFD (sound scene).

The experimental results from three test cases above indicates that
\begin{itemize}
\item Leveraging a pre-trained model (e.g., BEATs) for ESDD task is the most effective approach rather than training a model from scratch. 
\item By leveraging the pre-trained model, we can train model on AEFD (sound event) and then apply on the inference process for both AEFD (sound event) and ASDF (sound scene). 
This leads an advantage when currently published sound event datasets are larger and more diverse than sound scene datasets.
\end{itemize}



\begin{figure}[t]
    \centering
   \includegraphics[width=1.0\linewidth] {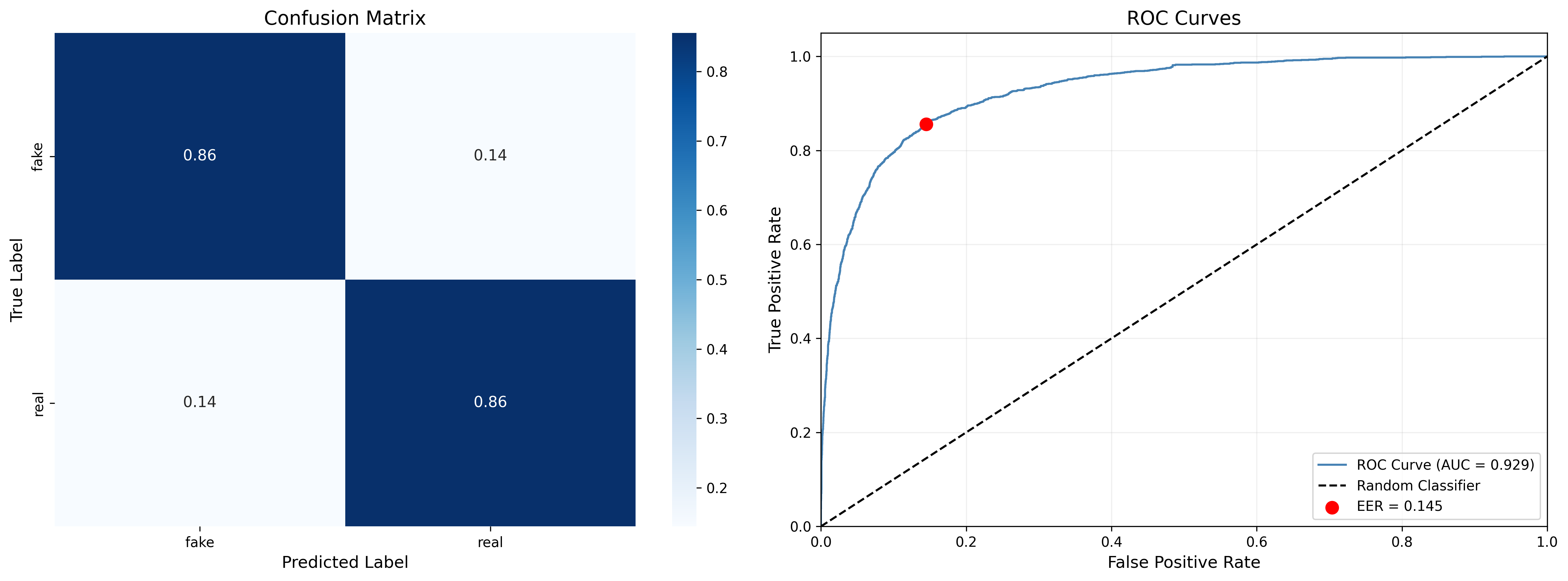}
    \caption{Confusions matrix result on ESDD-Challenge-TestSet (BEATs-finetune-MulHead model trained on sound events of EnvSDD training set)}    
    \label{fig:res_f1}
\end{figure}

\begin{figure}[t]
    \centering
   \includegraphics[width=1.0\linewidth] {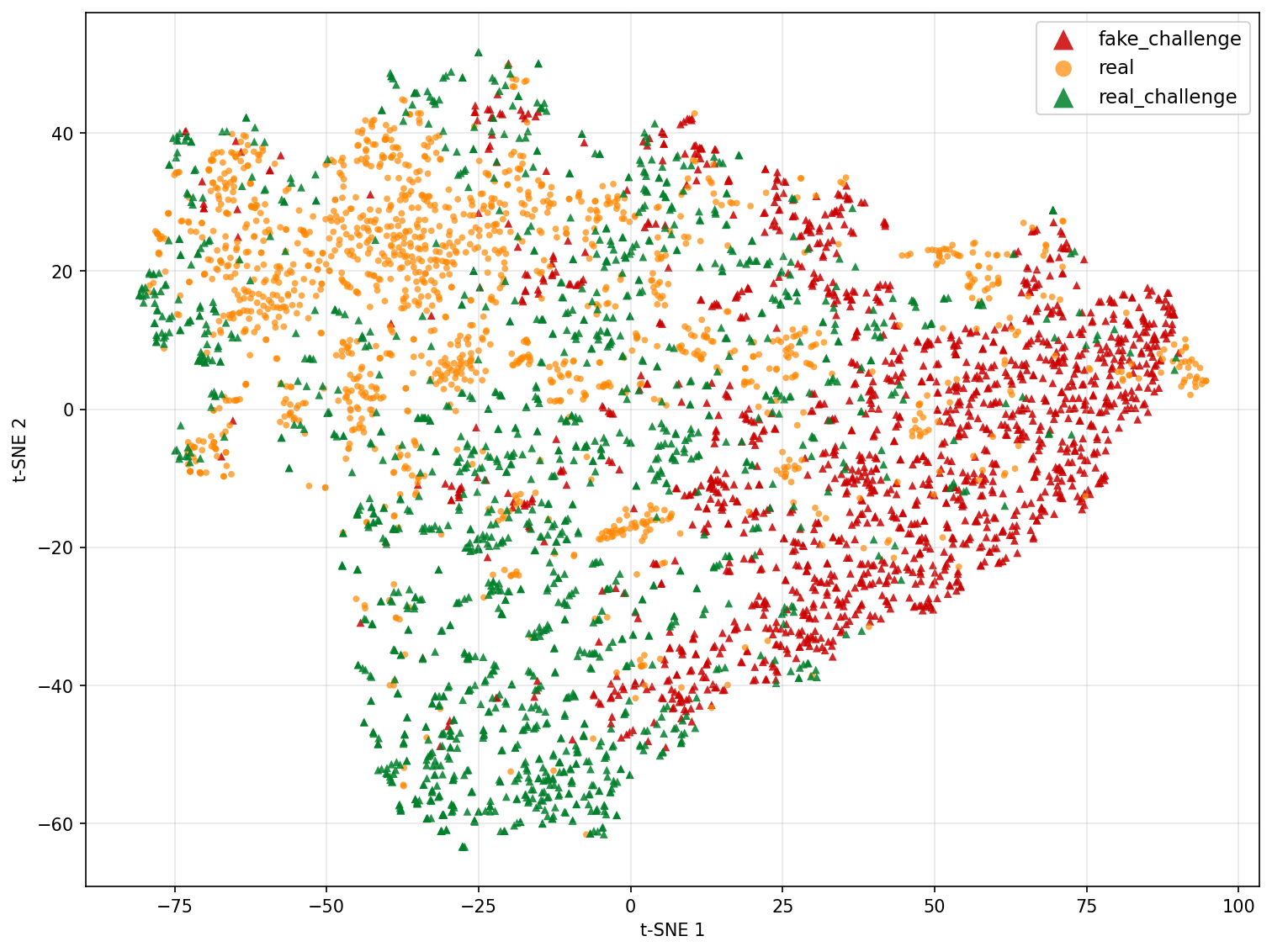}
    \caption{Distribution of real audio from EnvSDD, real and fake audio from ESDD-Challenge-TestSet}    
    \label{fig:res_f2}
\end{figure}

\textbf{Finetune the pre-trained BEATs model with proposed two-phase training strategy}:
As experimental results from three test cases shows above, it proves that leveraging a pre-trained model is more effective than training a model from scratch.
We therefore finetune the pre-trained BEATs model using our proposed two-phase training strategy presented in Section~\ref{sec_fine}.
This model is referred to as BEATs-finetune+MulHead.
The BEATs-finetune+MulHead is not only evaluated on EnvSDD dataset~\cite{data_01} but is also conducted the cross-dataset evaluation on ESDD-Challenge-TestSet~\cite{source_5}.

The experimental results in Table~\ref{tab:finetune01} shows that finetuning BEATs model is effective to enhance the ESDD performance compared with the embedding-based approach (BEATs-Emb+MLP).
We achieve the best model performance with Acc. of 0.98, F1 Score of 0.95, and AuC score of 0.99 on the Test subset of EnvSDD dataset (i.e., these results are computed on both sound event and sound scene of Test subset).

We then separate sound events and sound scenes in Test subset of EnvSDD dataset. Then, we plot the distribution of real and fake audio embeddings $/mathbf{X}$ using the TSNE algorithm.
As Fig.~\ref{fig:tsne_02} shows, fake and real audio scenes are well separated. Although unknown fake audio scenes (e.g., ATAAudioldm2, TTA-Audiolcm, TTA-Tangoflux) distribute in the middle space of real and seen fake audio scenes, unknown fake audio is not overlapped on real audio.
Regarding the distribution of fake and real audio events shown in Fig.~\ref{fig:tsne_01}, unknown fake audio events spreads widely and overlap on real audio events partly.

Given the experimental results from finetuning the pre-trained BEATs model with our proposed two-phase training strategy, we can conclude that
\begin{itemize}
    \item Detecting deepfake audio events is more challenging compared with deepfake audio scenes. It can be explained that audio events are more diverse.
    \item By focusing on distribution of real audio (Mahalanobis distribution) rather than the probability threshold, we avoid the incorrect detection regarding unseen fake audio which distributes in the boundary space between seen real and seen fake audio. 
    \item By evaluating the incorrect detection of deepfake sound event across resource data and fake generators in Fig.~\ref{fig:inc_01} and Fig.~\ref{fig:inc_02}, it indicates that most incorrect detection occurs on unknown fake data generated from real resource of Clotho, DCASE2023-Task7 or from the fake generator ATA-Audioldm2. This can be explained that these unknown fake data is not included in the training set.
\end{itemize}

\textbf{Compare with state-of-the-art models on EnvSDD dataset}:
To compare with the state-of-the-art models on EnvSDD dataset, we follow the setting mentioned in the original paper~\cite{data_01}, then separating the Test subset into four subsets (e.g., Test 01, Test 02, Test 03, Test 04). We then evaluate the proposed BEATs-finetune+MulHead model on these four subsets.
As experimental results show in Table~\ref{tab:per_cmp}, our proposed BEATs-finetune+MulHead model only trained on sound events outperforms the state-of-the-art models.
From the experimental results, we can conclude that:
\begin{itemize}
    \item Leverage a pre-trained model and train on sound events are effective to achieve a high-performance model for both sound event and sound scene deepfake detection.
    \item Finetuning a pre-trained model (BEATs), which was trained on sound events, is more effective rather than a pre-trained model trained on speech (W2V2+AASIST or AASIST)
\end{itemize}

\textbf{Cross-dataset Evaluation}: Given the high performance of the proposed BEATs-finetune+MulHead on EnvSDD dataset, we conduct the cross-dataset evaluation on ESDD-Challenge-TestSet~\cite{source_5} dataset.
In particular, we use the BEATs-finetune+MulHead model, which was trained on Training subset of EnvSDD dataset, to evaluate on ESDD-Challenge-TestSet~\cite{source_5} dataset.
As Table~\ref{tab:finetune02} shows, it again indicates that the model with training on sound event from EnvSDD dataset outperforms models with training on sound scene or both sound event and sound scene. 
In other words, training a model using sound scenes from EnvSDD dataset leads to downgrade the model performance.
This can be explained that audio data in ESDD-Challenge-TestSet is from VGG-Sound~\cite{source_6} which present audio events.
As Fig.~\ref{fig:res_f1} and Table~\ref{tab:finetune02} show, the high and balancing performance of Acc. score (0.86), F1 Score (0.80), and AuC score (0.93) from model only trained on sound event data prove that our proposed BEATs-finetune+Multihead model is robust and general.

We then plot the distribution of real audio from EnvSDD (e.g., training set) and the distribution of real/fake audio from ESDD-Challenge-TestSet (e.g., testing set) on Fig.~\ref{fig:res_f2}. The figure indicates that the fake audio of test set is well separated from the real audio of training set. 
As we use the distribution of real data from the training set to decide fake/real result of a unseen test data, it again proves that our proposed model is robust for cross-data evaluation.
However, the wide distribution of real audio from the ESDD-Challenge-TestSet (e.g., testing set) indicates that (1) real audio for sound event is very diverse and (2) model should be trained on as much diverse real audio resource as possible to make the distribution of real audio more condense.

\section{Conclusion}
We have presented a deep-learning based framework for Environmental Sound Deepfake Detection (ESDD). By conducting extensive experiments on the benchmark datasets of EnvSDD and ESDD-Challenge-TestSet datasets, we indicate that (1) Detecting deepfake audio of sound event and sound scene should be considered as individual tasks; (2) A model for ESDD task trained on sound event can be effectively applied for ESDD task on sound scene; (3) finetuning a pre-trained model is effective for ESDD task rather than training a model from scratch.
We eventually propose the best model for the ESDD task by combining finetuning the pre-trained BEATs model with our proposed two-phase training strategy.



\bibliographystyle{IEEEbib}
\bibliography{refs}

@inproceedings{data_01,
  title={EnvSDD: Benchmarking Environmental Sound Deepfake Detection},
  author={Yin, Han and Xiao, Yang and Das, Rohan Kumar and Bai, Jisheng and Liu, Haohe and Wang, Wenwu and Plumbley, Mark D},
  booktitle={Proc. INTERSPEECH},
  pages={201--205},
  year={2025}
}

@article{data_02,
  title={Scenefake: An initial dataset and benchmarks for scene fake audio detection},
  author={Yi, Jiangyan and Wang, Chenglong and Tao, Jianhua and Zhang, Chu Yuan and Fan, Cunhang and Tian, Zhengkun and Ma, Haoxin and Fu, Ruibo},
  journal={Pattern Recognition},
  volume={152},
  pages={110468},
  year={2024}
}

@misc{data_03,
  title = {Foley sound synthesis},
  note = {\url{https://dcase.community/challenge2023/task-foley-sound-synthesis}},
}

@misc{source_1,
  title = {DCASE-2019-Challenge-Task-1},
  howpublished = {\url{https://dcase.community/challenge2019/task-acoustic-scene-classification}},
  note = {Accessed: 2010-09-30}
}

@misc{dcase20213t7,
  title = {DCASE-2023-Challenge-Task-7},
  howpublished = {\url{https://dcase.community/challenge2023}},
  note = {Accessed: 2010-09-30}
}

@misc{source_2,
  title = {DCASE-2016-Challenge-Task-3},
  howpublished = {\url{https://dcase.community/challenge2016/task-sound-event-detection-in-real-life-audio}},
  note = {Accessed: 2010-09-30}
}

@misc{source_3,
  title = {DCASE-2017-Challenge-Task-3},
  howpublished = {\url{https://dcase.community/challenge2017/task-sound-event-detection-in-real-life-audio}},
  note = {Accessed: 2010-09-30}
}

@inproceedings{source_4,
    Author = {Salamon, J. and Jacoby, C. and Bello, J. P.},
    Booktitle = {Proc. ACM-MM},
    Pages = {1041--1044},
    Title = {A Dataset and Taxonomy for Urban Sound Research},
    Year = {2014}
}

@misc{source_5,
  title = {ESDD Challenge in ICASSP},
  howpublished = {\url{https://github.com/apple-yinhan/EnvSDD}},
  note = {Accessed: 2010-09-30}
}

@InProceedings{source_6,
  author       = "Honglie Chen and Weidi Xie and Andrea Vedaldi and Andrew Zisserman",
  title        = "VGGSound: A Large-scale Audio-Visual Dataset",
  booktitle    = "International Conference on Acoustics, Speech, and Signal Processing ",
  year         = "2020",
}

@inproceedings{novel_train,
  title={DIN-CTS: Low-complexity depthwise-inception neural network with contrastive training strategy for deepfake speech detection},
  author={Pham, Lam and Tran, Dat and Lam, Phat and Skopik, Florian and Schindler, Alexander and Poletti, Silvia and Fischinger, David and Boyer, Martin},
  booktitle={Proc. EUSIPCO},
  pages={556--560},
  year={2025}
  }

@inproceedings{paper_01,
  title={Detection of deepfake environmental audio},
  author={Ouajdi, Hafsa and Hadder, Oussama and Tailleur, Modan and Lagrange, Mathieu and Heller, Laurie M},
  booktitle={Proc. EUSIPCO},
  pages={196--200},
  year={2024}
}

@article{paper_02,
  title={Representation Loss Minimization with Randomized Selection Strategy for Efficient Environmental Fake Audio Detection},
  author={Phukan, Orchid Chetia and others},
  journal={arXiv preprint arXiv:2409.15767},
  year={2024}
}

@InProceedings{beats,
  title = 	 {{BEAT}s: Audio Pre-Training with Acoustic Tokenizers},
  author =       {Chen, Sanyuan and others},
  booktitle = 	 {Proceedings of the 40th International Conference on Machine Learning},
  pages = 	 {5178--5193},
  year = 	 {2023},
  volume = 	 {202}
}

@inproceedings{mixup1,
  title={Mixup-based acoustic scene classification using multi-channel convolutional neural network},
  author={Xu, Kele and Feng, Dawei and Mi, Haibo and Zhu, Boqing and Wang, Dezhi and Zhang, Lilun and Cai, Hengxing and Liu, Shuwen},
  booktitle={Pacific Rim Conference on Multimedia},
  pages={14-23},
  year={2018},
  url={https://doi.org/10.1007/978-3-030-00764-5\_2}
}

@inproceedings{mixup2,
  title={Learning from between-class examples for deep sound recognition},
  author={Tokozume, Yuji and Ushiku, Yoshitaka and Harada, Tatsuya},
  booktitle={ICLR},
  year={2018}
}

@inproceedings{lam_01,
  title={Bag-of-features models based on C-DNN network for acoustic scene classification},
  author={Pham, Lam and Yue, Lang and others},
  booktitle={Proc. AES},
  year={2019}
}

@article{lam_02,
  title={Lightweight deep neural networks for acoustic scene classification and an effective visualization for presenting sound scene contexts},
  author={Pham, Lam and Ngo, Dat and Salovic, Dusan and Jalali, Anahid and Schindler, Alexander and Nguyen, Phu X and Tran, Khoa and Vu, Hai Canh},
  journal={Applied Acoustics},
  volume={211},
  pages={109489},
  year={2023}
}

@inproceedings{lam_03,
  title={Wider or deeper neural network architecture for acoustic scene classification with mismatched recording devices},
  author={Pham, Lam and Tran, Khoa and Ngo, Dat and Tang, Hieu and Phan, Son and Schindler, Alexander},
  booktitle={Proceedings of the 4th ACM International Conference on Multimedia in Asia},
  pages={1--5},
  year={2022}
}

@inproceedings{lam_04,
  title={Multi-view audio and music classification},
  author={Phan, Huy and Le Nguyen, Huy and Ch{\'e}n, Oliver Y and Pham, Lam and Koch, Philipp and McLoughlin, Ian and Mertins, Alfred},
  booktitle={Proc. ICASSP},
  pages={611--615},
  year={2021}
}

@article{Adam,
  title={Adam: A Method for Stochastic Optimization},
  author={Diederik,  P. K. and Jimmy,  B.},
  journal={CoRR},
  year={2015},
  volume={abs/1412.6980}
}

@inproceedings{audioset,
title	= {Audio Set: An ontology and human-labeled dataset for audio events},
author	= {Jort F. Gemmeke and Daniel P. W. Ellis and Dylan Freedman and Aren Jansen and Wade Lawrence and R. Channing Moore and Manoj Plakal and Marvin Ritter},
booktitle	= {Proc. ICASSP},
year	= {2017}
}

@INPROCEEDINGS{data_clotho,
  author={Drossos, Konstantinos and Lipping, Samuel and Virtanen, Tuomas},
  booktitle={Proc. ICASSP}, 
  title={Clotho: an Audio Captioning Dataset}, 
  year={2020},
  pages={736-740}
}

@inproceedings{envsdd_challenge,
  title={Environmental sound deepfake detection challenge: An overview},
  author={Yin, Han and Xiao, Yang and Das, Rohan Kumar and Bai, Jisheng and Dang, Ting},
  booktitle={Proc. ICASSP},
  pages={21772--21774},
  year={2026}
  }

@article{wavlm,
  title={Wavlm: Large-scale self-supervised pre-training for full stack speech processing},
  author={Chen, Sanyuan and Wang, Chengyi and Chen, Zhengyang and Wu, Yu and Liu, Shujie and Chen, Zhuo and Li, Jinyu and Kanda, Naoyuki and Yoshioka, Takuya and Xiao, Xiong and others},
  journal={IEEE Journal of Selected Topics in Signal Processing},
  volume={16},
  number={6},
  pages={1505--1518},
  year={2022}
}

@inproceedings{MultiLibrispeech, 
   title={MLS: A Large-Scale Multilingual Dataset for Speech Research},
   booktitle={Proc. INTERSPEECH},
   author={Pratap, Vineel and Xu, Qiantong and Sriram, Anuroop and Synnaeve, Gabriel and Collobert, Ronan},
   year={2020},
   pages={2757–2761}
   }

@inproceedings{whisper,
  title={Robust speech recognition via large-scale weak supervision},
  author={Radford, Alec and others},
  booktitle={Proc. ICML},
  pages={28492--28518},
  year={2023}
}

@inproceedings{wav2vec20,
  author={Alexis Conneau and Alexei Baevski and Ronan Collobert and Abdelrahman Mohamed and Michael Auli},
  title={{Unsupervised Cross-Lingual Representation Learning for Speech Recognition}},
  year=2021,
  booktitle={Proc. INTERSPEECH},
  pages={2426--2430}
}

@article{Contrastive_loss,
  title={SINCERE: Supervised Information Noise-Contrastive Estimation REvisited},
  author={Feeney, Patrick and Hughes, Michael C},
  journal={arXiv preprint arXiv:2309.14277},
  year={2023}
}
\end{document}